\documentclass[journal]{IEEEtran}
\usepackage[T1]{fontenc}
\usepackage{ifpdf}
\usepackage{cite}
\usepackage[pdftex]{graphicx}
\usepackage{amsmath}
\usepackage{amssymb}
\usepackage{bm}
\usepackage{array}
\usepackage{fixltx2e}

\usepackage{balance} 

\usepackage{mathtools} 

\usepackage{diagbox}

\usepackage{makecell}

\usepackage{threeparttable}

\usepackage{color,soul}
\soulregister\cite7

\usepackage{algorithm}
\usepackage{algpseudocode}

\usepackage{setspace}

\usepackage{amsthm}

\newtheorem{theorem}{Theorem}
\newtheorem{proposition}{Proposition}
 
\newtheorem{lemma}{Lemma}

\newtheorem{definition}{Definition}

\makeatletter
\newcommand{\doublewidetilde}[1]{{%
  \mathpalette\double@widetilde{#1}%
}}
\newcommand{\double@widetilde}[2]{%
  \sbox\z@{$\m@th#1\widetilde{#2}$}%
  \ht\z@=.85\ht\z@
  \widetilde{\box\z@}%
}
\makeatother

\usepackage{bbm}

\usepackage{subcaption}

\usepackage[hidelinks]{hyperref}

\usepackage{bookmark}

\begin{document}

\title{\vspace{-5mm} \Large Joint Power Allocation and Reflecting-Element Activation for Energy Efficiency Maximization in IRS-Aided Communications Under CSI Uncertainty}

\author{\normalsize Christos N. Efrem and Ioannis Krikidis, \IEEEmembership{Fellow, IEEE} \vspace{-7mm} 
\thanks{This work was supported by the Cyprus Research and Innovation Foundation under Grant DUAL USE/0922/0031 (RISE), and by the 
European Union's Horizon Europe programme through the European Research Council (ERC) under Grant Agreement No.~101112697 (WAVE). 

The authors are with the Department of Electrical and Computer Engineering, University of Cyprus, 1678 Nicosia, Cyprus (e-mail:  \{efrem.christos, krikidis\}@ucy.ac.cy). }}

\markboth{T\MakeLowercase{his article has been accepted for publication in} IEEE C\MakeLowercase{ommunications} L\MakeLowercase{etters}, D\MakeLowercase{ecember} 2025. $\copyright$2025 IEEE.}%
{}

\maketitle

\begin{abstract}

We study the joint power allocation and reflecting-element (RE) activation to maximize the energy efficiency (EE) in communication systems assisted by an intelligent reflecting surface (IRS), taking into account imperfections in channel state information (CSI). The robust optimization problem is mixed-integer, i.e., the optimization variables are continuous (transmit power) and discrete (binary states of REs). In order to solve this challenging problem we develop two algorithms. The first one is an alternating optimization (AO) method that attains a suboptimal solution with low complexity, based on the Lambert $W$ function and a dynamic programming (DP) algorithm. The second one is a branch-and-bound (B\&B) method that uses AO as its subroutine and is formally guaranteed to achieve a globally optimal solution. Both algorithms do not require any external optimization solver for their implementation. Furthermore, numerical results show that the proposed algorithms outperform the baseline schemes, AO achieves near-optimal performance in most cases, and B\&B has low computational complexity on average.     

\end{abstract}


\vspace{-2mm}
\section{Introduction}

Intelligent reflecting surfaces (IRSs) can dynamically reconfigure the phase of the incident electromagnetic waves by using adjustable electronic circuits \cite{Liaskos2018}. The optimization of~EE in IRS-assisted wireless networks has been an important area of research \cite{Huang2019}. Furthermore, there has been increased interest in IRS beamforming to maximize the signal-to-noise ratio (SNR)~\cite{Zhang2022} and the sum rate \cite{Zhao2021}. Moreover, the joint optimization of IRS location and number of REs has been examined in \cite{Efrem2023} from the outage-probability perspective. The optimal number of REs that maximizes the spectral efficiency and EE, considering channel estimation and feedback, has been studied in \cite{Zappone2021}. In addition, the authors in \cite{Khaleel2022} have investigated a low-cost strategy for only tuning the on/off states of REs, while keeping their phase shifts fixed. A comparison between IRS and decode-and-forward relaying has been presented in \cite{Bjornson2020}.

This letter extends our previous work \cite{Efrem2024} on the robust optimization of EE under CSI uncertainty. In particular, the focus now is on the joint optimization of the REs' on/off states \emph{and} the transmit power. The considered problem becomes more complicated due to the mixed nature of optimization variables: the transmit power is continuous, while the states of REs are discrete. Firstly, we design an AO method based on the Lambert $W$ function and the DP algorithm given in~\cite{Efrem2024}. Also, we develop a B\&B method that is based on the AO algorithm. \emph{Interestingly, the convergence of B\&B to an optimal solution is guaranteed provided that the AO initialization point is appropriately chosen.} In this way, we can evaluate the AO performance by comparing with the global optimum of B\&B.

The rest of this letter is structured as follows. Section \ref{section:System_Model} presents the system model and formulates the robust optimization problem. Next, Sections \ref{section:Alternating Optimization} and \ref{section:Global Optimization_BnB} develop and analyze the AO and B\&B algorithms, respectively. Finally, Section \ref{section:Numerical_Results} gives numerical results, while Section \ref{section:Conclusion} concludes the letter.

The mathematical notation is the same as \cite[Section I-D]{Efrem2024}. We write $x \coloneqq y$, or $y \eqqcolon x$, whenever $x$ is by definition equal to $y$. In addition, for every integer $k$, $W_k(z)$ denotes the $k^{\text{th}}$ complex branch of the Lambert $W$ function. 




\vspace{-2mm}
\section{System Model and Problem Formulation} \label{section:System_Model}

We consider a single-antenna transmitter (Tx) that communicates with a single-antenna receiver (Rx) via a passive IRS with $N$ REs, denoted by $\mathcal{N} \coloneqq \{1,\dots,N\}$. Let $h_0 \in \mathbb{C}$ be the Tx-Rx direct channel, and $h_n \in \mathbb{C}$ be the cascaded channel corresponding to the $n^\text{th}$ RE whose phase shift is denoted by $\phi_n \in [0,2\pi)$ \cite[Section~II-A]{Efrem2024}.\footnote{The channel coefficients are fixed in a given time frame (flat fading). Also, the number of bits used for controlling the IRS phase shifts is assumed to be sufficiently large, so the phase shifts are (approximately) continuous.} For convenience, we define $\mathbf{h} = [h_0,\dots,h_N]^\top \in \mathbb{C}^{N+1}$ and express each channel in polar coordinates as ${h_n} = {\alpha_n} {e^{j\theta_n}}$, where $\alpha_n = | h_n | \geq 0$ and $\theta_n = \operatorname{Arg}(h_n) \in [0,2\pi)$, for all $n \in \mathcal{N}_0 \coloneqq \{0,\dots,N\}$. Regarding the CSI, we model the actual channel $\mathbf{h}$ as the sum of two terms: the estimated channel $\widehat{\mathbf{h}} = [\widehat{h}_0,\dots,\widehat{h}_N]^\top \in \mathbb{C}^{N+1}$ (after quantization), and the unknown estimation error $\widetilde{\mathbf{h}} = [\widetilde{h}_0,\dots,\widetilde{h}_N]^\top \in \mathbb{C}^{N+1}$, i.e., $\mathbf{h} = \widehat{\mathbf{h}} + \widetilde{\mathbf{h}}$. The elements of $\widehat{\mathbf{h}}$ and $\widetilde{\mathbf{h}}$ can also be expressed in polar coordinates: ${\widehat{h}_n} = {\widehat{\alpha}_n} {e^{j\widehat{\theta}_n}}$ and ${\widetilde{h}_n} = {\widetilde{\alpha}_n} {e^{j\widetilde{\theta}_n}}$. 

Furthermore, we adopt a deterministic CSI-error model: ${\lVert \widetilde{\mathbf{h}} \rVert}_2 \leq \xi$, where $\xi \in [0,\widehat{\alpha}_{\min}]$ is the CSI-uncertainty radius and $\widehat{\alpha}_{\min} \coloneqq \min_{n \in \mathcal{N}_0} \{ \widehat{\alpha}_n \}$; see \cite[Section~II-B]{Efrem2024} for more details.\footnote{In practice, given the channel-estimation and quantization procedures, we can determine the parameters $\xi$ and $q_{\min}$ (the minimum quantization level for channel magnitudes, which is independent of $N$). If we ensure that $\xi \leq q_{\min}$, then we have $\xi \leq \widehat{\alpha}_{\min}$ since $q_{\min} \leq \widehat{\alpha}_{\min}$. This condition may not apply in cases with a very large number of REs due to increased $\xi$. Nevertheless, a moderate number of REs is usually preferable, since more REs (despite the SNR improvement) result in higher energy consumption and channel-estimation overhead.} Because the estimation error $\widetilde{\mathbf{h}}$ is unknown, the IRS phase shifts are selected using only the estimated channel $\widehat{\mathbf{h}}$ so that the ideal SNR (with $\mathbf{h} = \widehat{\mathbf{h}}$) is maximized, i.e., $\phi_n =  (\widehat{\theta}_0 - \widehat{\theta}_n) \bmod {2\pi}$, for all $n \in \mathcal{N}$ \cite[Eq.~(6)]{Efrem2024}. In addition, every RE is either on/activated (operating in reflection mode) or off/deactivated (operating in absorption mode) \cite{Zhao2021,Khaleel2022}. For this reason, we use a binary vector ${{\mathbf{x}}} = {[{x_1}, \ldots ,{x_N}]^\top \in \{0,1\}^N}$, where $x_n = 1$ if and only if the $n^\text{th}$ RE is on. Given the CSI uncertainty, the worst-case SNR is given by \cite[Theorem~3]{Efrem2024} 
\begin{equation}  \label{equation:Worst-case_SNR} 
{\gamma}_{\textnormal{w}} (p,\mathbf{x};\xi) =  \frac{p}{\sigma^2} \left( f(\mathbf{x}) - g(\mathbf{x};\xi) \right)^2,
\end{equation}
where $p \geq 0$ is the transmit power, ${\sigma^2}>0$ is the noise power, $f(\mathbf{x}) =  \widehat{\alpha}_0 + \sum_{n \in \mathcal{N}} {x_n \widehat{\alpha}_n}$, and $g(\mathbf{x};\xi) = \xi \sqrt{1 + \sum_{n \in \mathcal{N}} {x_n}}$. Note that $f(\mathbf{x}) \geq g(\mathbf{x};\xi)$, for all $\mathbf{x} \in \{0,1\}^N$.

Afterwards, the worst-case energy efficiency is defined by 
\begin{equation} \label{equation:Worst-case_EE}
{\operatorname{EE}_{\text{w}}}(p,\mathbf{x};\xi) = \frac{\log_2 \left( 1 + {\gamma_{\text{w}}} (p,\mathbf{x};\xi) \right)}{\operatorname{P}_\text{tot} (p,\mathbf{x})}  ,
\end{equation} 
where $\operatorname{P}_\text{tot} (p,\mathbf{x})$ is the total power consumption. In particular, 
$\operatorname{P}_\text{tot} (p,\mathbf{x}) =  \eta^{-1} p + (P_{\text{on}} - P_{\text{off}}) {\sum_{n \in \mathcal{N}} {x_n}} + P_{\text{fix}}$, with $P_{\text{fix}} = P_{\text{static}} + N P_{\text{off}}$, where $\eta \in (0,1]$ is the power amplifier's efficiency, and $P_{\text{static}} > 0$ accounts for the dissipated power in the remaining signal-processing blocks at the transmitter and receiver. In addition, $P_{\text{on}}$ and $P_{\text{off}}$ (with $P_{\text{on}} \geq P_{\text{off}} > 0$) represent the power consumption of each activated and deactivated RE, respectively \cite{Efrem2024}. Note that $P_{\text{fix}}$ is defined \emph{differently} from \cite[Eq.~(30)]{Efrem2024}: $P_{\text{fix}}^{\text{old}} = \eta^{-1} p + P_{\text{static}}$.

Now, we formulate a robust optimization problem as follows 
\begin{subequations} \label{problem:EE_original}
\begin{alignat}{3}
 \operatorname{EE}_{\text{w}}^{*} \! \coloneqq & \mathop {\text{max}} \limits_{p,\mathbf{x}} & \quad & {\operatorname{EE}_{\text{w}}}(p,\mathbf{x};\xi)   \\
  & \,\:\text{s.t.} & & {\gamma_{\text{w}}} (p,\mathbf{x};\xi) \geq \gamma_{\min} ,  \\
  & & & 0 \leq p \leq p_{\max}  , \;  \mathbf{x} \in \{ 0,1 \}^N ,  
\end{alignat}
\end{subequations}
where $\gamma_{\min} > 0$ is the minimum required SNR, and $p_{\max} > 0$ \linebreak is the maximum transmit power. Specifically, we want to maximize the worst-case EE by jointly adjusting the transmit power and tuning the on/off states of REs, while satisfying a minimum-SNR requirement and a maximum-power constraint. Problem~\eqref{problem:EE_original} is a challenging \emph{mixed-integer optimization problem}, i.e., with continuous and discrete variables.\footnote{In \cite{Efrem2024} we assume fixed transmit power, thus having only discrete variables (the REs' on/off states) without the maximum-power constraint.}  

For convenience in the design of B\&B algorithm (see Section~\ref{section:Global Optimization_BnB}), we define the following problem  
\begin{subequations} \label{subproblem:EE} 
\begin{alignat}{3}
 \operatorname{EE}_{\text{w}}^{**} (p_\ell,p_u) \! \coloneqq & \mathop {\text{max}} \limits_{p,\mathbf{x}} & \quad & {\operatorname{EE}_{\text{w}}}(p,\mathbf{x};\xi)   \\
  & \,\:\text{s.t.} & & {\gamma_{\text{w}}} (p,\mathbf{x};\xi) \geq \gamma_{\min} ,  \\
  & & & p_\ell \leq p \leq p_u  , \; \mathbf{x} \in \{ 0,1 \}^N ,   
\end{alignat}
\end{subequations}
which is a generalization of problem \eqref{problem:EE_original}, obtained by setting $p_\ell = 0$ and $p_u = p_{\max}$. In general $0 \leq p_\ell \leq p_u \leq p_{\max}$, so every feasible solution to problem~\eqref{subproblem:EE} is also feasible for problem \eqref{problem:EE_original}. The following proposition gives a necessary and sufficient condition for feasibility.

\begin{proposition} \label{proposition:Feasibility_condition}
Problem~\eqref{subproblem:EE} is feasible if and only if ${\gamma}_{\textnormal{w}} (p_u,\mathbf{1}_N;\xi) \geq \gamma_{\min}$.  
\end{proposition}

\begin{IEEEproof}
It is sufficient to show that ${\partial {\gamma_{\textnormal{w}}} (p,\mathbf{x};\xi)} / {\partial p} \geq 0$ and ${\partial {\gamma_{\textnormal{w}}} (p,\mathbf{x};\xi)} / {\partial x_n} \geq 0$, $\forall n \in \mathcal{N}$. From \eqref{equation:Worst-case_SNR} we deduce that ${\partial {\gamma_{\textnormal{w}}} (p,\mathbf{x};\xi)} / {\partial p} = {\left( f(\mathbf{x}) - g(\mathbf{x};\xi) \right)^2} / {\sigma^2} \geq 0$. In addition, the monotonicity of ${\gamma_{\textnormal{w}}} (p,\mathbf{x};\xi)$ with respect to each $x_n$ follows from \cite[Proposition~5]{Efrem2024}.  
\end{IEEEproof}

\vspace{-2mm}
\section{Alternating Optimization Algorithm} \label{section:Alternating Optimization}

In this section, we will design an AO algorithm in order to achieve a suboptimal solution to problem \eqref{subproblem:EE}; the original problem \eqref{problem:EE_original} is just a special case.\footnote{More specifically, we start with a feasible solution, and then we alternately optimize a subset of variables with the remaining variables being fixed.}  

\subsection{Transmit Power Allocation with Fixed REs' On/Off States}

Given the vector $\mathbf{x} \in \{ 0,1 \}^N$, problem \eqref{subproblem:EE} reduces to
\begin{subequations} \label{problem:EE_fixed_on/off_states}
\begin{alignat}{3}
  & \mathop {\text{max}} \limits_{p} & \quad & {\operatorname{EE}_{\text{w}}}(p,\mathbf{x};\xi)\\
  & \,\:\text{s.t.} & & {\gamma_{\text{w}}} (p,\mathbf{x};\xi) \geq \gamma_{\min} ,  \; p_{\ell} \leq p \leq p_u  .    
\end{alignat}
\end{subequations}
Note that $\left( f(\mathbf{x}) - g(\mathbf{x};\xi) \right)^2 > 0$, because $\gamma_{\min} > 0$ $\implies$ ${\gamma_{\text{w}}} (p,\mathbf{x};\xi) > 0$. For convenience, let us define the functions
\begin{equation}
u(\mathbf{x};\xi) =  {\left( f(\mathbf{x}) - g(\mathbf{x};\xi) \right)^2}/{\sigma^2}  ,
\end{equation}
\begin{equation}
v(\mathbf{x}) = (P_{\text{on}} - P_{\text{off}}) {\sum_{n \in \mathcal{N}} {x_n}} + P_{\text{fix}}  .
\end{equation}
Observe that $u(\mathbf{x};\xi), v(\mathbf{x}) > 0$. Problem \eqref{problem:EE_fixed_on/off_states} is equivalent to 
\begin{subequations} \label{problem:EE_fixed_on/off_states_equivalent}
\begin{alignat}{3}
  & \mathop {\text{max}} \limits_{p} & \quad & {\operatorname{EE}_{\text{w}}}(p,\mathbf{x};\xi)\\
  & \,\:\text{s.t.} & & p'_{\ell} \leq p \leq p_u , 
\end{alignat} 
\end{subequations}
where $p'_{\ell} = \max \left( \frac{\gamma_{\min}}{u(\mathbf{x};\xi)},p_{\ell} \right)$.

\begin{proposition} \label{proposition:Optimal_transmit_power}
The optimal solution to problem \eqref{problem:EE_fixed_on/off_states}/\eqref{problem:EE_fixed_on/off_states_equivalent} is   
\begin{equation} \label{equation:p_opt}
p_{\textnormal{opt}} = \min\left(\max\left(p'_{\ell},\widetilde{p}_{\textnormal{opt}}\right),p_u\right) ,
\end{equation}
where $\widetilde{p}_{\textnormal{opt}} = ( e^{{W_0 \left( \left( u(\mathbf{x};\xi) v(\mathbf{x}) \eta - 1 \right) e^{-1} \right)}+1} - 1 )/{u(\mathbf{x};\xi)}$. 
\end{proposition}

\begin{IEEEproof}
The worst-case EE in \eqref{equation:Worst-case_EE} can be expressed as ${\operatorname{EE}_{\text{w}}}(p,\mathbf{x};\xi) = \frac{\log_2 \left( 1 + u(\mathbf{x};\xi) p \right)}{\eta^{-1} p + v(\mathbf{x})}$. By differentiating with respect to $p$, we obtain 
\begin{equation}
\tfrac{\partial}{\partial p} {\operatorname{EE}_{\text{w}}}(p,\mathbf{x};\xi) = \tfrac{s(p,\mathbf{x};\xi)}{\log(2) \left( 1 + u(\mathbf{x};\xi) p \right) \left( \eta^{-1} p + v(\mathbf{x}) \right)^2} ,
\end{equation}
where $s(p,\mathbf{x};\xi) = u(\mathbf{x};\xi) \left( \eta^{-1} p + v(\mathbf{x}) \right) - \eta^{-1} \left( 1 + u(\mathbf{x};\xi) p \right) \log \left( 1 + u(\mathbf{x};\xi) p \right)$. To find the optimal (unconstrained) transmit power $p \geq 0$, we should solve the equation $\tfrac{\partial}{\partial p} {\operatorname{EE}_{\text{w}}}(p,\mathbf{x};\xi) = 0$ which yields $s(p,\mathbf{x};\xi) = 0$, i.e., 
\begin{equation*} 
\frac{1 + u(\mathbf{x};\xi) p}{e} \log \left( \frac{1 + u(\mathbf{x};\xi) p}{e} \right)  =  \frac{u(\mathbf{x};\xi) v(\mathbf{x}) \eta - 1}{e}  .
\end{equation*}
By applying the transformation $y = \log \left( \frac{1 + u(\mathbf{x};\xi) p}{e} \right) \iff p = (e^{y+1} - 1)/{u(\mathbf{x};\xi)}$, the above equation becomes $y e^y = \left( u(\mathbf{x};\xi) v(\mathbf{x}) \eta - 1 \right) e^{-1}$. 

\begin{lemma}[\hspace{1sp}\cite{Knuth1996}]
The equation $y e^y = \delta$, where $\delta$ is a real number, has the following real solution(s)
\begin{equation}
y = \left\{ 
\begin{array}{ll}
W_0(\delta)\ \textnormal{or}\ W_{-1}(\delta) , & \textnormal{if}\,\  -e^{-1} \leq \delta < 0 ,  \\
W_0(\delta) , & \textnormal{if}\,\  \delta \geq 0 . \\
\end{array}
\right.
\end{equation} 
If $\delta < -e^{-1}$, then the equation has no real solution. There is a unique real solution $y  = W_0(-e^{-1}) = W_{-1}(-e^{-1}) = -1$ when $\delta = -e^{-1}$, whereas there are exactly two (distinct) real solutions when $-e^{-1} < \delta < 0$ (because $W_0(\delta) > -1 > W_{-1}(\delta)$ in this region). Also, $W_0(\delta) \geq 0$ when $\delta \geq 0$. 
\end{lemma}

In our case, $u(\mathbf{x};\xi) p \geq 0 \implies y \geq -1$, so there is a unique solution given by $y = W_0 \left( \left( u(\mathbf{x};\xi) v(\mathbf{x}) \eta - 1 \right) e^{-1} \right) > -1$. Hence, the optimal (unconstrained) transmit power, $\widetilde{p}_{\textnormal{opt}} > 0$, is expressed as shown in Proposition~\ref{proposition:Optimal_transmit_power}. In particular, it holds that $\tfrac{\partial^2}{\partial p^2} {\operatorname{EE}_{\text{w}}}(\widetilde{p}_{\text{opt}},\mathbf{x};\xi) < 0$, because $\tfrac{\partial}{\partial p} s(\widetilde{p}_{\text{opt}},\mathbf{x};\xi) = - \eta^{-1} u(\mathbf{x};\xi) \log \left( 1 + u(\mathbf{x};\xi) \widetilde{p}_{\text{opt}} \right) < 0$. Therefore, $\widetilde{p}_{\text{opt}}$ is a local maximum that is also the unique global maximum of ${\operatorname{EE}_{\text{w}}}(p,\mathbf{x};\xi)$ for $p \geq 0$, since ${\operatorname{EE}_{\text{w}}}(0,\mathbf{x};\xi) = 0$ and $\lim_{p \to \infty} {\operatorname{EE}_{\text{w}}}(p,\mathbf{x};\xi) = 0$. Finally, the (unique) optimal solution to the constrained problem \eqref{problem:EE_fixed_on/off_states_equivalent} can be easily computed by \eqref{equation:p_opt}, because  ${\operatorname{EE}_{\text{w}}}(p,\mathbf{x};\xi)$ is increasing for $p \in [0,\widetilde{p}_{\textnormal{opt}})$ and decreasing for $p \in (\widetilde{p}_{\textnormal{opt}},\infty)$. 
\end{IEEEproof}

\subsection{Reflecting-Element Activation with Fixed Transmit Power}

Given the transmit power $p \in [p_\ell,p_u]$, problem \eqref{subproblem:EE} becomes
\begin{subequations} \label{problem:EE_fixed_p}
\begin{alignat}{3}
  & \mathop {\text{max}} \limits_{\mathbf{x}} & \quad & {\operatorname{EE}_{\text{w}}}(p,\mathbf{x};\xi)   \\
  & \,\:\text{s.t.} & & {\gamma_{\text{w}}} (p,\mathbf{x};\xi) \geq \gamma_{\min} ,  \; \mathbf{x} \in \{ 0,1 \}^N .  
\end{alignat}
\end{subequations}
Problem \eqref{problem:EE_fixed_p} is a discrete (binary) optimization problem that can be globally solved using the dynamic programming algorithm given in \cite[Algorithm~1]{Efrem2024}, which has polynomial complexity $O(N {\log N})$; please refer to \cite[Theorem~4]{Efrem2024}.

\subsection{Algorithm Analysis}

The AO procedure is presented in Algorithm \ref{algorithm:AO}. First, the algorithm decides the feasibility of problem \eqref{subproblem:EE} (step 1) and initializes some parameters (step 2). Subsequently, it computes two solutions in an alternating manner by first optimizing with respect to: i) the transmit power $p$ (steps 3--7), and ii) the binary vector $\mathbf{x}$ (steps 8--13). Finally, it returns the best of them (steps 14--15).\footnote{In addition to the $\epsilon$-convergence criterion, we can also use a (predetermined) maximum number of iterations for robustness against numerical errors.} The following theorem demonstrates the convergence and complexity of the AO algorithm.

\noindent
\begin{minipage}[!t]{\columnwidth}
\begin{algorithm}[H]   
\caption{Alternating Optimization (AO) for problem~\eqref{subproblem:EE}}  \label{algorithm:AO}
\footnotesize  
\begin{algorithmic}[1] 
\State \textbf{if} {${\gamma}_{\textnormal{w}} (p_u,\mathbf{1}_N;\xi) < \gamma_{\min}$} \textbf{then return} \textit{`Infeasible'} \textbf{end if} 
	 
\State Choose a feasible solution $(p_{\text{opt}}^{(0)},\mathbf{x}_{\text{opt}}^{(0)})$ to problem~\eqref{subproblem:EE}, and a  
\Statex \, convergence tolerance $\epsilon > 0$. ${\operatorname{EE}_{\text{w}}^{(0)}} \leftarrow {\operatorname{EE}_{\text{w}}}(p_{\text{opt}}^{(0)},\mathbf{x}_{\text{opt}}^{(0)};\xi)$, $i \leftarrow 0$     

\Repeat 
	\State $i \leftarrow i+1$, $p_{\text{opt}}^{(i)} \leftarrow p_{\text{opt}}$ according to \eqref{equation:p_opt} with $\mathbf{x} = \mathbf{x}_{\text{opt}}^{(i-1)}$   
	\State Solve problem \eqref{problem:EE_fixed_p} with $p = p_{\text{opt}}^{(i)}$, using \cite[Algorithm~1]{Efrem2024}, and let 
	\Statex \quad\, $\mathbf{x}_{\text{opt}}^{(i)}$ be its globally optimal solution. ${\operatorname{EE}_{\text{w}}^{(i)}} \leftarrow {\operatorname{EE}_{\text{w}}}(p_{\text{opt}}^{(i)},\mathbf{x}_{\text{opt}}^{(i)};\xi)$       
\Until{$| {\operatorname{EE}_{\text{w}}^{(i)}} - {\operatorname{EE}_{\text{w}}^{(i-1)}} | < \epsilon$} 
\State $(p'_{\text{opt}},\mathbf{x}'_{\text{opt}}) \leftarrow (p_{\text{opt}}^{(i)},\mathbf{x}_{\text{opt}}^{(i)})$, $\operatorname{EE}'_{\text{w}} \leftarrow \operatorname{EE}_{\text{w}}^{(i)}$, $i \leftarrow 0$

\Repeat 
	\State $i \leftarrow i+1$. Solve problem \eqref{problem:EE_fixed_p} with $p = p_{\text{opt}}^{(i-1)}$, using   
	\Statex \quad\, \cite[Algorithm~1]{Efrem2024}, and let $\mathbf{x}_{\text{opt}}^{(i)}$ be its globally optimal solution. 
	\State $p_{\text{opt}}^{(i)} \leftarrow p_{\text{opt}}$ according to \eqref{equation:p_opt} with $\mathbf{x} = \mathbf{x}_{\text{opt}}^{(i)}$    
	\State ${\operatorname{EE}_{\text{w}}^{(i)}} \leftarrow {\operatorname{EE}_{\text{w}}}(p_{\text{opt}}^{(i)},\mathbf{x}_{\text{opt}}^{(i)};\xi)$  
\Until{$| {\operatorname{EE}_{\text{w}}^{(i)}} - {\operatorname{EE}_{\text{w}}^{(i-1)}} | < \epsilon$} 
\State $(p''_{\text{opt}},\mathbf{x}''_{\text{opt}}) \leftarrow (p_{\text{opt}}^{(i)},\mathbf{x}_{\text{opt}}^{(i)})$, $\operatorname{EE}''_{\text{w}} \leftarrow \operatorname{EE}_{\text{w}}^{(i)}$

\State \textbf{if} {$\operatorname{EE}'_{\text{w}} \geq \operatorname{EE}''_{\text{w}}$} \textbf{then return} $(p'_{\text{opt}},\mathbf{x}'_{\text{opt}})$ and $\operatorname{EE}'_{\text{w}}$  
\State \textbf{else} \textbf{return} $(p''_{\text{opt}},\mathbf{x}''_{\text{opt}})$ and $\operatorname{EE}''_{\text{w}}$ \textbf{end if} 
\end{algorithmic}
\end{algorithm} 
\end{minipage} 
\vspace{1mm}

\begin{theorem} \label{theorem:Alternating_Optimization_algorithm}
Algorithm~\ref{algorithm:AO} returns either \textit{`Infeasible'} if problem \eqref{subproblem:EE} is not feasible, or a feasible solution otherwise. In the latter case, the algorithm produces a nondecreasing sequence of objective values in each loop, i.e., ${\operatorname{EE}_{\text{w}}^{(i)}} \geq {\operatorname{EE}_{\text{w}}^{(i-1)}}$ for all $i \geq 1$, and terminates in finitely many iterations. Moreover, its complexity is $O(I N {\log N})$, where $I$ is the total number of iterations and $N$ is the number of reflecting elements. 
\end{theorem}

\begin{IEEEproof}
It can be easily seen that Algorithm~\ref{algorithm:AO} correctly returns \textit{`Infeasible'} due to Proposition~\ref{proposition:Feasibility_condition}. Moreover, we can show by induction on $i$ that $(p_{\text{opt}}^{(i)},\mathbf{x}_{\text{opt}}^{(i)})$ is a feasible solution to problem~\eqref{subproblem:EE}, for all iterations $i \geq 0$. Also, we have 
\begin{equation}
\scalebox{0.95}{$
\begin{split}
{\operatorname{EE}_{\text{w}}^{(i-1)}} & \coloneqq {\operatorname{EE}_{\text{w}}}(p_{\text{opt}}^{(i-1)},\mathbf{x}_{\text{opt}}^{(i-1)};\xi) \leq {\operatorname{EE}_{\text{w}}}(p_{\text{opt}}^{(j)},\mathbf{x}_{\text{opt}}^{(k)};\xi) \\
& \leq {\operatorname{EE}_{\text{w}}}(p_{\text{opt}}^{(i)},\mathbf{x}_{\text{opt}}^{(i)};\xi) \eqqcolon {\operatorname{EE}_{\text{w}}^{(i)}} , \quad \forall i \geq 1  , 
\end{split} $}
\end{equation} 
where $(j,k) = (i,i-1)$ and $(j,k) = (i-1,i)$ in the first and second loop, respectively. As a result, the sequence $\{ {\operatorname{EE}_{\text{w}}^{(i)}} \}_{i \geq 0}$ is nondecreasing and, since it is upper bounded (i.e., ${\operatorname{EE}_{\text{w}}}(p,\mathbf{x};\xi) \leq P_{\text{fix}}^{-1} \log_2 \left( 1 + {\gamma_{\text{w}}} (p_{\max},\mathbf{1}_N;\xi) \right) < \infty$), it converges to a finite value. Therefore,  $\lim_{i \to \infty} (\operatorname{EE}_{\text{w}}^{(i)} - {\operatorname{EE}_{\text{w}}^{(i-1)}}) = 0$, which by the limit definition means that: for every $\epsilon > 0$, there exists an integer $m$ such that if $i \geq m$ then $| {\operatorname{EE}_{\text{w}}^{(i)} - {\operatorname{EE}_{\text{w}}^{(i-1)}}} | < \epsilon$. Thus, Algorithm \ref{algorithm:AO} terminates in a finite number of iterations.  

Finally, the complexity of Algorithm 1 is $O(I N {\log N})$, because each iteration requires $O(N {\log N})$ time, mainly due to \cite[Algorithm~1]{Efrem2024} in steps 5 and 9. 
\end{IEEEproof}

\vspace{-2mm}
\section{Branch-and-Bound Method} \label{section:Global Optimization_BnB}

A powerful approach for global optimization is the B\&B technique. 
The main idea is to develop a B\&B method based on the AO algorithm to achieve fast convergence to an optimal solution. In a nutshell, B\&B generates multiple subproblems by recursively splitting the feasible set and using bounds on their optimum values. 
In our case, we have the original problem~\eqref{problem:EE_original} and subproblems in the form of~\eqref{subproblem:EE}. A generated subproblem is called \emph{active} if it has not been examined yet. 

Moreover, $\mathcal{Q}$ is the first-in-first-out list (i.e., a queue) that contains the active subproblems. For each subproblem~\eqref{subproblem:EE} we just store the interval $[p_{\ell},p_u]$ in $\mathcal{Q}$. Initially the list contains the original problem~\eqref{problem:EE_original}, i.e., $\mathcal{Q} = \{ [0,p_{\max}] \}$, whereas $\mathcal{Q} = \varnothing$ in the last iteration. We also denote its cardinality by $Q = |\mathcal{Q}|$, and the maximum $Q$ over all iterations by $Q_{\max}$ (which is proportional to the space complexity of B\&B). In addition, $(\widetilde{p},\widetilde{\mathbf{x}})$ is the best feasible solution found so far, and $\widetilde{\operatorname{EE}}_{\text{w}}$ is the current best energy efficiency, i.e., $\widetilde{\operatorname{EE}}_{\text{w}} = {\operatorname{EE}_{\text{w}}}(\widetilde{p},\widetilde{\mathbf{x}};\xi)$. Note that B\&B produces a \emph{nondecreasing} sequence of $\widetilde{\operatorname{EE}}_{\text{w}}$ values over its iterations. Finally, $\overline{\operatorname{EE}}_{\text{w}} = \overline{\operatorname{EE}}_{\text{w}} (p_{\ell},p_u)$ and $\underline{\operatorname{EE}}_{\text{w}} = \underline{\operatorname{EE}}_{\text{w}} (p_{\ell},p_u)$ are upper and lower bounds (U/LB) on the optimum value of the subproblem, respectively, i.e., $\underline{\operatorname{EE}}_{\text{w}} (p_{\ell},p_u) \leq \operatorname{EE}_{\text{w}}^{**} (p_\ell,p_u) \leq \overline{\operatorname{EE}}_{\text{w}} (p_\ell,p_u)$.

In particular, the UB is chosen as follows 
\begin{subequations} \label{subproblem:Upper_Bound}
\begin{alignat}{3}
 \overline{\operatorname{EE}}_{\text{w}} (p_\ell,p_u) \! \coloneqq & \mathop {\text{max}} \limits_{\mathbf{x}} & \quad & \frac{\log_2 \left( 1 + {\gamma_{\text{w}}} (p_u,\mathbf{x};\xi) \right)}{\operatorname{P}_\text{tot} (p_\ell,\mathbf{x})}   \\
  & \,\:\text{s.t.} & & {\gamma_{\text{w}}} (p_u,\mathbf{x};\xi) \geq \gamma_{\min} ,  \\
  & & & \mathbf{x} \in \{ 0,1 \}^N . 
\end{alignat}
\end{subequations}
It holds that $\operatorname{EE}_{\text{w}}^{**} (p_\ell,p_u) \leq \overline{\operatorname{EE}}_{\text{w}} (p_\ell,p_u)$, because the objective function of \eqref{subproblem:EE} is upper bounded by the objective function of \eqref{subproblem:Upper_Bound}, and the feasible set of \eqref{subproblem:EE} is a subset of the feasible set of \eqref{subproblem:Upper_Bound}. This problem can be (globally) solved using again \cite[Algorithm~1]{Efrem2024} with $P_{\text{fix}}^{\text{old}} = \eta^{-1} p_\ell  + P_{\text{static}}$ and $\overline{\gamma} = p_u / \sigma^2$. 

Regarding the LB, $\underline{\operatorname{EE}}_{\text{w}} (p_{\ell},p_u)$ is computed by Algorithm~\ref{algorithm:AO} provided that subproblem \eqref{subproblem:EE} is feasible. In particular, Algorithm~\ref{algorithm:AO} is initialized with $p_{\text{opt}}^{(0)} = p_u$; this is \emph{very important for the B\&B convergence} (see the proof of Theorem~\ref{theorem:BnB_algorithm}).

The proposed B\&B is given in Algorithm~\ref{algorithm:BnB}.\footnote{With `\textbf{continue}' the program skips any remaining statements in the while-loop for the current iteration, and continues from the next iteration.} Specifically, it performs the following basic operations: 1) \emph{subproblem selection} from the list $\mathcal{Q}$, 2) \emph{infeasibility}: remove a subproblem when it is not feasible, 3) \emph{bounding}: compute upper and lower bounds on the optimum value of the subproblem, 4) \emph{update} of $\widetilde{\operatorname{EE}}_{\text{w}}$ and $(\widetilde{p},\widetilde{\mathbf{x}})$, 5) \emph{pruning/fathoming}: remove a subproblem when $\overline{\operatorname{EE}}_{\text{w}} \leq \widetilde{\operatorname{EE}}_{\text{w}} + \varepsilon$ (the subproblem does not contain any better solution with respect to the given tolerance $\varepsilon > 0$), or when $\underline{\operatorname{EE}}_{\text{w}} = \overline{\operatorname{EE}}_{\text{w}}$ (we have already found an optimal solution to the subproblem), and 6) \emph{branching}: split the feasible set into two subsets using the standard bisection technique, i.e., $[p_{\ell},p_u] = [p_{\ell},p_m] \cup [p_m,p_u]$, where $p_m = \frac{1}{2}(p_\ell + p_u)$.      


\noindent
\begin{minipage}[!t]{\columnwidth}
\begin{algorithm}[H]   
\caption{Branch-and-Bound (B\&B) for problem~\eqref{problem:EE_original}}  \label{algorithm:BnB}
\footnotesize  
\begin{algorithmic}[1] 

\State Choose an optimal-solution accuracy $\varepsilon > 0$.   
\State $\mathcal{Q} \leftarrow \{ [0,p_{\max}] \}$, $\widetilde{\operatorname{EE}}_{\text{w}} \leftarrow -\infty$, $i \leftarrow 0$ 

\While{$\mathcal{Q} \neq \varnothing$} 
	\State $i \leftarrow i+1$ 
	\State $\triangleright$ \textit{Subproblem selection}
	\State Let $[p_\ell,p_u]$ be the first subproblem in the front of the list $\mathcal{Q}$. 
	\State $\mathcal{Q} \leftarrow \mathcal{Q} \setminus \{[p_\ell,p_u]\}$ 
	
	\State $\triangleright$ \textit{Pruning due to infeasibility} 
	\State \textbf{if} {${\gamma}_{\textnormal{w}} (p_u,\mathbf{1}_N;\xi) < \gamma_{\min}$}  \textbf{then continue end if} 
	
	\State $\triangleright$ \textit{Bounding} 
	\State Compute the bound $\overline{\operatorname{EE}}_{\text{w}}$ by solving problem~\eqref{subproblem:Upper_Bound}, using 
	\Statex \quad\, \cite[Algorithm~1]{Efrem2024} with $P_{\text{fix}}^{\text{old}} = \eta^{-1} p_\ell  + P_{\text{static}}$ and $\overline{\gamma} = p_u / \sigma^2$. 
	\State Compute a feasible solution $(p,\mathbf{x})$ to subproblem~\eqref{subproblem:EE}, using   
	\Statex \quad\, Algorithm \ref{algorithm:AO} with $(p_{\text{opt}}^{(0)},\mathbf{x}_{\text{opt}}^{(0)}) = (p_u,\mathbf{1}_N)$, and let 
	\Statex \quad\, $\underline{\operatorname{EE}}_{\text{w}} = {\operatorname{EE}_{\text{w}}}(p,\mathbf{x};\xi)$ be its objective value. 
	
	\State $\triangleright$ \textit{Updating $\widetilde{\operatorname{EE}}_{\text{w}}$ and $(\widetilde{p},\widetilde{\mathbf{x}})$} 
	\State \textbf{if} {$\underline{\operatorname{EE}}_{\text{w}} > \widetilde{\operatorname{EE}}_{\text{w}}$} \textbf{then} $\widetilde{\operatorname{EE}}_{\text{w}} \leftarrow \underline{\operatorname{EE}}_{\text{w}}$, $(\widetilde{p},\widetilde{\mathbf{x}}) \leftarrow (p,\mathbf{x})$ \textbf{end if} 
	
	\State $\triangleright$ \textit{Pruning/Fathoming} 
	\State \textbf{if} {$(\overline{\operatorname{EE}}_{\text{w}} \leq \widetilde{\operatorname{EE}}_{\text{w}} + \varepsilon) \lor (\underline{\operatorname{EE}}_{\text{w}} = \overline{\operatorname{EE}}_{\text{w}})$} \textbf{then continue end if}  
	
	\State $\triangleright$ \textit{Branching and adding the new subproblems at the back of $\mathcal{Q}$} 
	\State $p_m \leftarrow \frac{1}{2}(p_\ell + p_u)$, $\mathcal{Q} \leftarrow \mathcal{Q} \cup \{[p_\ell,p_m],[p_m,p_u]\}$       
\EndWhile  

\State \textbf{if} {$\widetilde{\operatorname{EE}}_{\text{w}} = - \infty$} \textbf{then return} \textit{`Infeasible'} 
\State \textbf{else} \textbf{return} $(\widetilde{p},\widetilde{\mathbf{x}})$ and $\widetilde{\operatorname{EE}}_{\text{w}}$ \textbf{end if} 
\end{algorithmic}
\end{algorithm} 
\end{minipage}
\vspace{1mm}

In general, the B\&B method has exponential complexity in the worst case, but much lower (possibly polynomial) complexity on average. Since the computation of an exact optimal solution is generally impossible due to finite precision, we need a definition of approximately-optimal solutions. 

\begin{definition}[$\varepsilon$-optimal solution]
Consider the following optimization problem: $\varphi^{*} \coloneqq \max \, \{\varphi(\mathbf{y}): \mathbf{y} \in \mathcal{D}\}$, where $\varphi(\mathbf{y})$ is a continuous function and $\mathcal{D} \subset \mathbb{R}^n$ is a compact set. Given an $\varepsilon >0$, we say that $\widetilde{\mathbf{y}} \in \mathcal{D}$ is a (globally) \emph{$\varepsilon$-optimal solution} to the aforementioned problem if $\varphi^{*} - \varepsilon \leq \varphi(\widetilde{\mathbf{y}}) \left(\leq \varphi^{*}\right)$.  
\end{definition}

Now, we can prove the correctness of the B\&B algorithm.

\begin{theorem} \label{theorem:BnB_algorithm}
If the original problem~\eqref{problem:EE_original} is not feasible, then Algorithm~\ref{algorithm:BnB} returns `Infeasible' after a single iteration. Otherwise, it terminates within a finite number of iterations and $(\widetilde{p},\widetilde{\mathbf{x}})$ is an $\varepsilon$-optimal solution to problem~\eqref{problem:EE_original} with $\operatorname{EE}_{\text{w}}^{*} - \varepsilon \leq \widetilde{\operatorname{EE}}_{\text{w}} = {\operatorname{EE}_{\text{w}}}(\widetilde{p},\widetilde{\mathbf{x}};\xi) \leq \operatorname{EE}_{\text{w}}^{*}$.  
\end{theorem}

\begin{IEEEproof}
See the Appendix.   
\end{IEEEproof}

\vspace{-2mm}
\section{Numerical Results} \label{section:Numerical_Results}

The simulation parameters are similar to \cite[Section~VI]{Efrem2024} with the following differences: transmitter, receiver and IRS locations $(0,0,0)$, $(80,0,0)$ and $(40,10,5)$, respectively, channel Rician factors $\kappa_u = \kappa_v = 6\ \text{dB}$, $\xi = \tau {\widehat{\alpha}_{\min}}$, where $\tau \in [0,1]$, $\gamma_{\min} = {\chi} {\gamma}_{\text{w}} (p_{\max},\mathbf{1}_N;\widehat{\alpha}_{\min})$, where $\chi \in [0,1]$ (the feasibility of the original problem~\eqref{problem:EE_original} is ensured, because ${\gamma}_{\text{w}} (p_{\max},\mathbf{1}_N;\xi) \geq {\gamma}_{\text{w}} (p_{\max},\mathbf{1}_N;\widehat{\alpha}_{\min}) \geq \gamma_{\min}$), $N = 50$, $p_{\max} = 27\ \text{dBm}$, $\sigma^2 = -85\ \text{dBm}$, $P_\text{off} = 0.4\ \text{mW}$, $\chi = 0.4$.  In addition, we select $(p_{\text{opt}}^{(0)},\mathbf{x}_{\text{opt}}^{(0)}) = (p_u,\mathbf{1}_N)$ and $\epsilon = \varepsilon = 10^{-3}$.

For comparison purposes, we consider three baseline schemes: 1) \emph{Only Reflecting-Element Optimization (OREO)} with maximum transmit power, using \cite[Algorithm~1]{Efrem2024} with $p = p_{\max}$, 2) \emph{Only Power Allocation (OPA)} with all reflecting elements being activated, according to \eqref{equation:p_opt} with $(p_{\ell}, p_u, \mathbf{x}) = (0, p_{\max}, \mathbf{1}_N)$, and 3) \emph{Maximum Power and All-Reflecting-Elements Activation (MPAREA)}, i.e., $p = p_{\max}$ and $\mathbf{x} = \mathbf{1}_N$.

\begin{figure}[!t]
\centering
\includegraphics[width=0.78\linewidth]{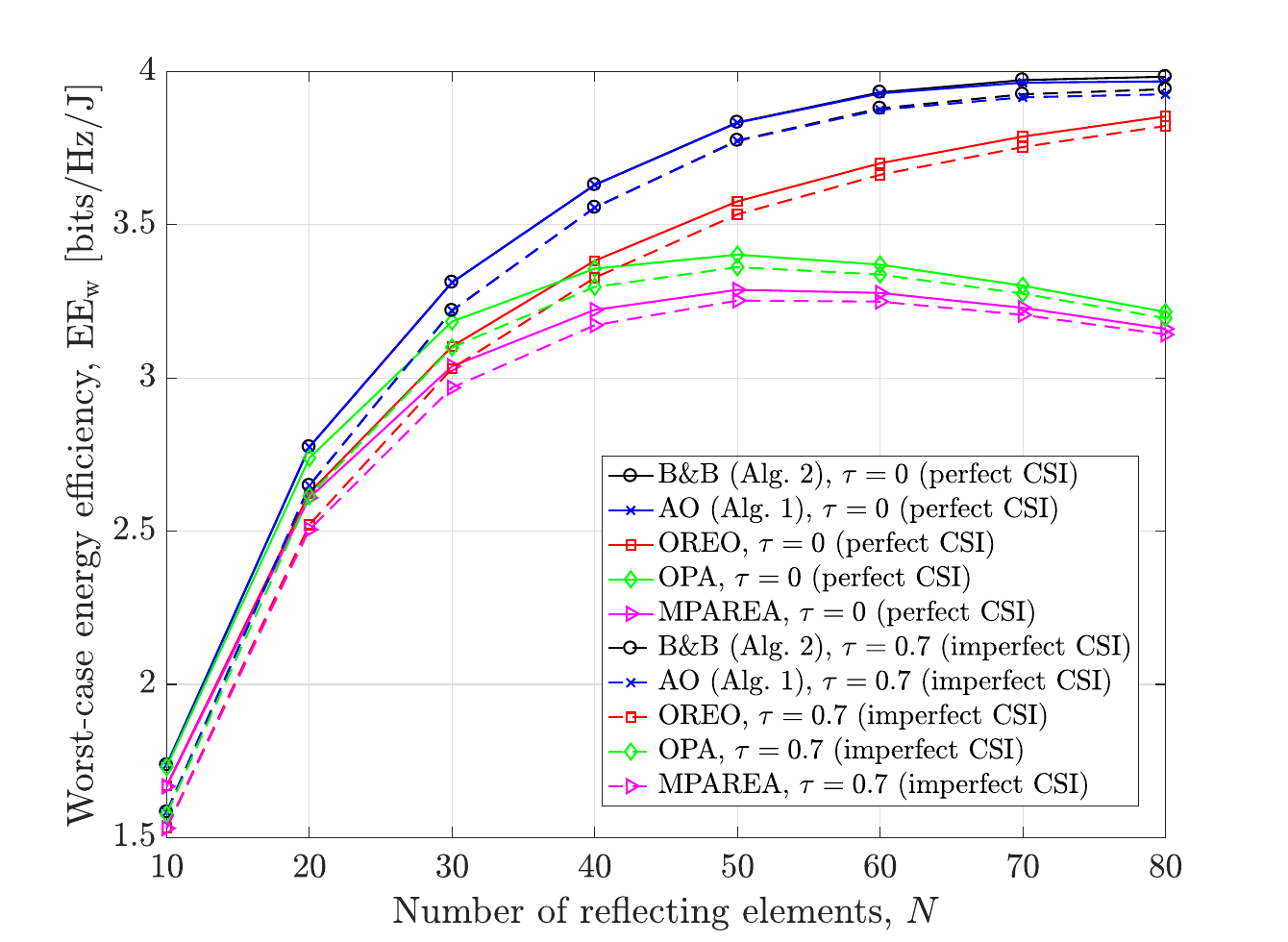}
\caption{Worst-case energy efficiency versus the number of reflecting elements, with CSI-uncertainty radius $\xi = \tau {\widehat{\alpha}_{\min}}$.}
\label{figure:EE_vs_N}
\vspace{-2mm}
\end{figure}

Fig.~\ref{figure:EE_vs_N} shows the worst-case EE against the number of REs. Firstly, we can observe that severe CSI uncertainty results in EE reduction for all schemes. Secondly, AO achieves nearly the same performance with B\&B, i.e., it is very close to the global optimum. Moreover, AO outperforms all benchmarks, with MPAREA having the worst performance. Thirdly, OPA is better than OREO for small $N$, whereas OREO shows higher performance than OPA for large $N$. Finally, it is interesting to note that the average running time was of the order of $10^{-3}\,\mathrm{s}$ for B\&B, $10^{-4}\,\mathrm{s}$ for AO, and $10^{-5}\,\mathrm{s}$ for all the benchmarks. AO required only 4--6 iterations (2--3 iterations for each repeat-until loop) to converge on average. Also, the average number of B\&B iterations was 143--584, and the average maximum number of active subproblems in the list $\mathcal{Q}$ ($Q_{\max}$) was 32--164  (i.e., low memory requirement or space complexity).

\begin{figure}[!t]
\centering
\includegraphics[width=0.73\linewidth]{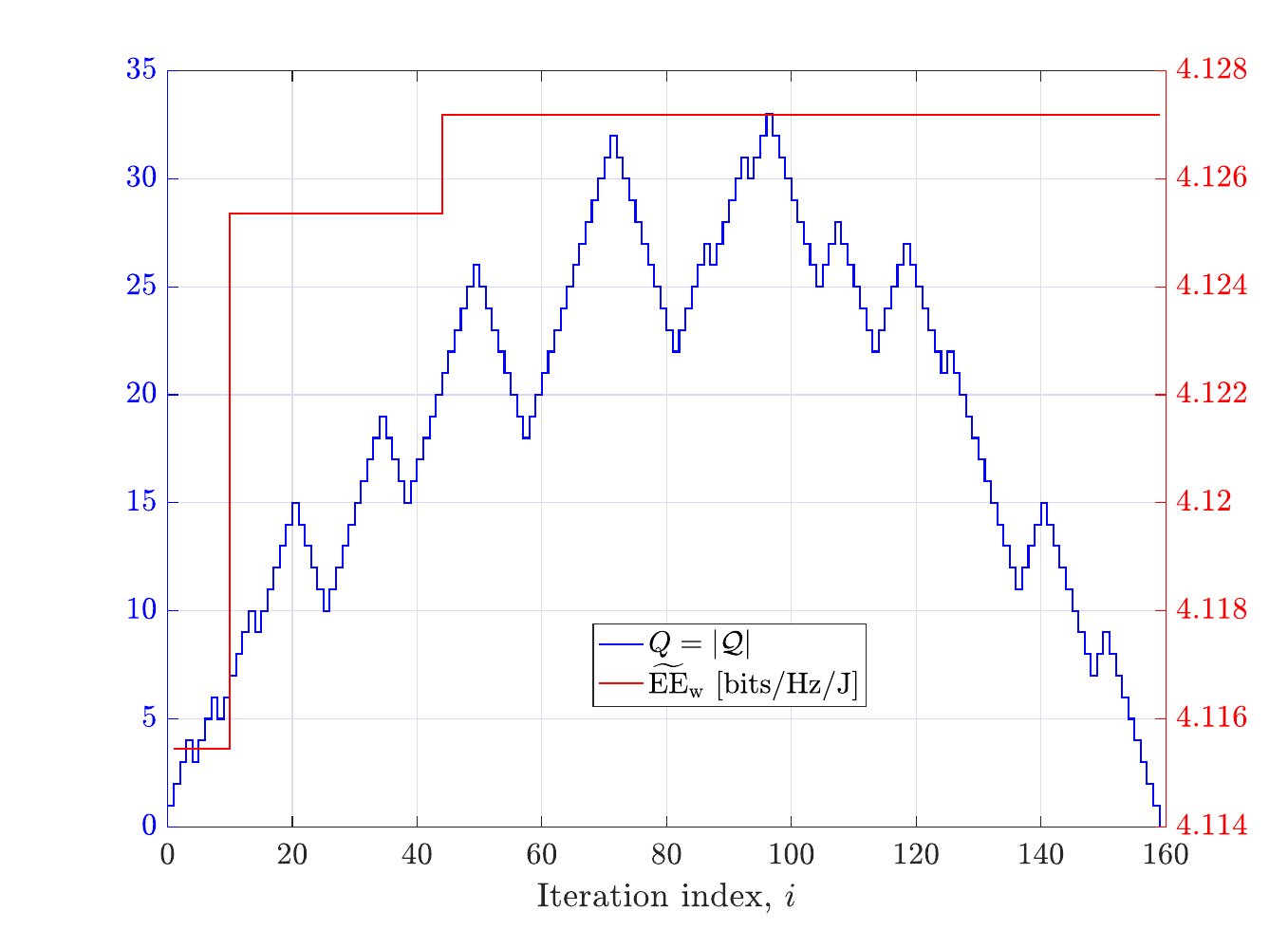}
\caption{B\&B evolution for a particular problem instance, with CSI-uncertainty radius $\xi = \tau {\widehat{\alpha}_{\min}}$ and $\tau = 0.7$.}
\label{figure:BnB_evolution}
\vspace{-2mm}
\end{figure}

Furthermore, the evolution of the B\&B algorithm for a problem instance is illustrated in Fig.~\ref{figure:BnB_evolution}. In particular, the total number of B\&B iterations is 159, while $Q_{\max} = 33$. We can also observe that: a) $Q=1$ at the beginning, whereas $Q=0$ at the end, and b) $\widetilde{\operatorname{EE}}_{\text{w}}$ is nondecreasing with iterations.



\begin{figure}[!t]
\centering
\includegraphics[width=0.75\linewidth]{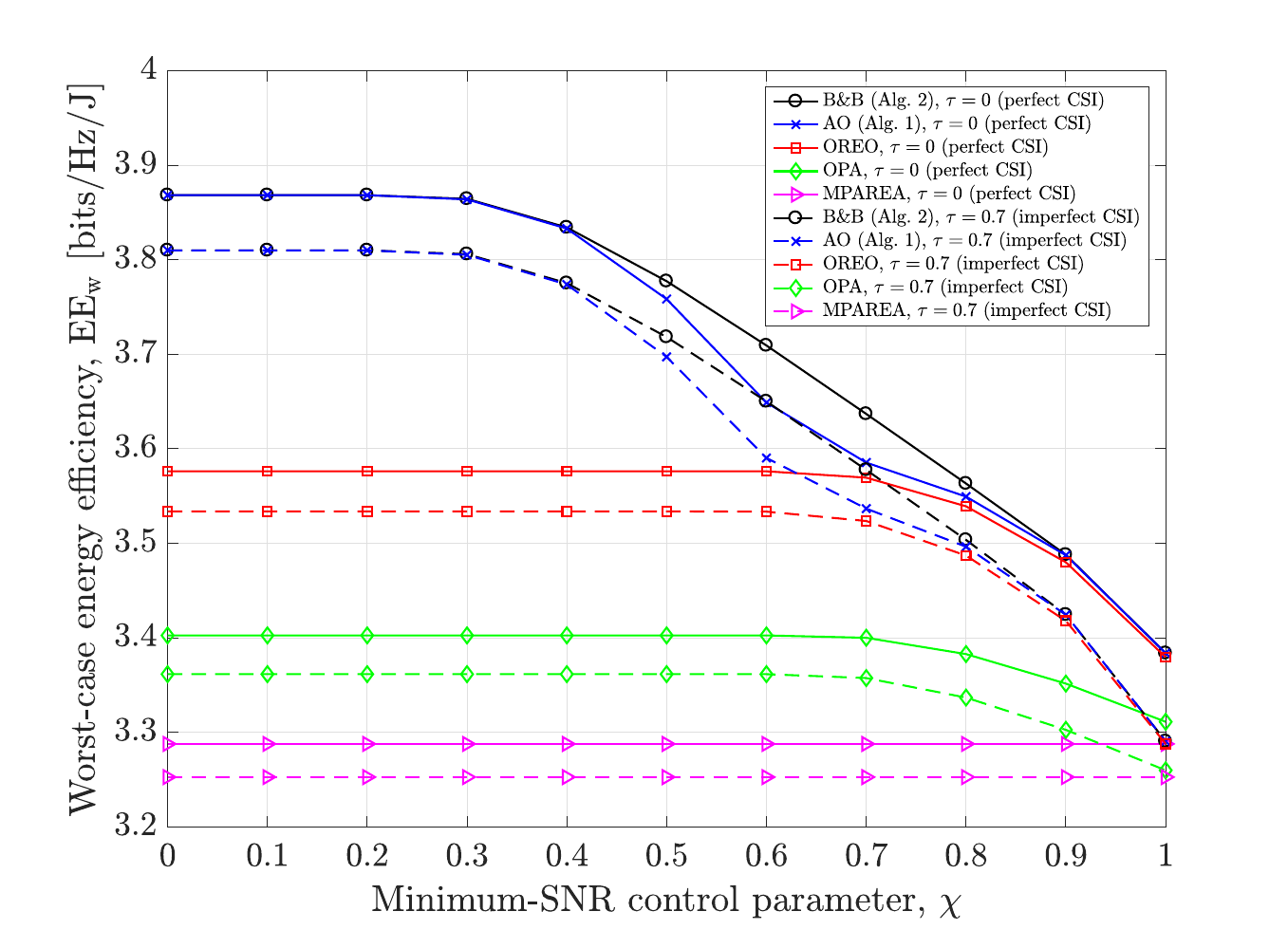}
\caption{Worst-case energy efficiency versus the minimum-SNR control parameter, with $\gamma_{\min} = {\chi} {\gamma}_{\text{w}} (p_{\max},\mathbf{1}_N;\widehat{\alpha}_{\min})$.} 
\label{figure:EE_vs_chi}
\vspace{-2mm}
\end{figure}

Finally, we investigate the impact of the minimum SNR on the EE. Based on Fig.~\ref{figure:EE_vs_chi}, the worst-case EE is a nonincreasing function of $\chi$ for all algorithms, since the increase of $\chi$ results in a more restricted feasible set. Although AO outperforms the baseline schemes and remains close to the global optimum in the majority of cases, there is a \emph{noticeable gap} between AO and B\&B in some cases (e.g., for $\chi = 0.5$\,--\,$0.8$).

\vspace{-2mm}
\section{Conclusion} \label{section:Conclusion}

In this letter, we dealt with the robust optimization of EE in IRS-aided communication systems under CSI uncertainty. In particular, we designed two algorithms (AO and B\&B) to jointly optimize the transmit power and the on/off states of REs. Numerical simulations verified the theoretical results and showed the effectiveness of the proposed algorithms.

\vspace{-2mm}
\appendix[Proof of Theorem~\ref{theorem:BnB_algorithm}] \label{appendix:Proof_of_theorem}

The first part of the theorem is easy to prove by observing that the original problem will be pruned due to infeasibility in the first iteration, thus  $\widetilde{\operatorname{EE}}_{\text{w}}$ will remain equal to $-\infty$. For the second part about the B\&B convergence to an $\varepsilon$-optimal solution in finite time, it is sufficient to show that 
\begin{equation} \label{equation:BnB_convergence_sufficient_condition}
\lim_{\Delta p \to 0} \left[ \overline{\operatorname{EE}}_{\text{w}} (p_{\ell},p_u) - \underline{\operatorname{EE}}_{\text{w}} (p_{\ell},p_u) \right] = 0 ,
\end{equation} 
where $\Delta p = p_u - p_\ell$ is the length of the interval $[p_\ell,p_u]$.\footnote{In Algorithm~\ref{algorithm:BnB}, immediately after step 14 we have $\widetilde{\operatorname{EE}}_{\text{w}} \geq \underline{\operatorname{EE}}_{\text{w}}$, thus $\overline{\operatorname{EE}}_{\text{w}} - \widetilde{\operatorname{EE}}_{\text{w}} \leq  \overline{\operatorname{EE}}_{\text{w}} - \underline{\operatorname{EE}}_{\text{w}}$. Therefore, condition~\eqref{equation:BnB_convergence_sufficient_condition} implies that there is a finite number of iterations after which $\overline{\operatorname{EE}}_{\text{w}} - \widetilde{\operatorname{EE}}_{\text{w}} \leq \varepsilon$. Eventually all subproblems will be pruned, resulting in $\mathcal{Q} = \varnothing$ and $\operatorname{EE}_{\text{w}}^{*} \leq \widetilde{\operatorname{EE}}_{\text{w}} + \varepsilon$.}

For convenience, using $p_\ell = p_u - \Delta p$ and assuming (without loss of generality) that $p_u$ is fixed, but arbitrary, we define 
\begin{equation}
F(\Delta p; \mathbf{x}) = \frac{\log_2 \left( 1 + {\gamma_{\text{w}}} (p_u,\mathbf{x};\xi) \right)}{\operatorname{P}_\text{tot} (p_u - \Delta p,\mathbf{x})}  .
\end{equation}
The first-order derivative of $F(\Delta p; \mathbf{x})$ with respect to $\Delta p$ is 
\begin{equation} 
\frac{\partial F(\Delta p; \mathbf{x})}{\partial (\Delta p)} =  \frac{\eta^{-1} \log_2 \left( 1 + {\gamma_{\text{w}}} (p_u,\mathbf{x};\xi) \right)}{\operatorname{P}_\text{tot}^2 (p_u - \Delta p,\mathbf{x})}   . 
\end{equation}
Observe that the derivative is \emph{continuous} for $\Delta p \leq p_u$, and can be bounded by a positive constant $M$, i.e., $| \partial F(\Delta p; \mathbf{x}) / \partial (\Delta p) | \leq M$, given by  
\begin{equation}
M = \eta^{-1} P_{\text{fix}}^{-2} \log_2 \left( 1 + {\gamma_{\text{w}}} (p_{\max},\mathbf{1}_N;\xi) \right) < \infty.
\end{equation}

Next, according to Taylor's theorem, we can express $F(\Delta p; \mathbf{x})$ around $\Delta p = 0$ as follows
\begin{equation} \label{equation:Taylor_formula}
F(\Delta p; \mathbf{x}) = F(0; \mathbf{x}) + R(\Delta p; \mathbf{x})  ,
\end{equation}
where $R(\Delta p; \mathbf{x}) = \int_0^{\Delta p} \frac{\partial F(t; \mathbf{x})}{\partial t} \mathrm{d} t$ is the remainder term, which can be bounded by
\begin{equation} \label{equation:Remainder_bound}
|R(\Delta p; \mathbf{x})| \leq M |\Delta p|  , \quad \forall \mathbf{x} \in \{0,1\}^N .
\end{equation} 

Now, we define the  set $\mathcal{X} = \{\mathbf{x} \in \{0,1\}^N : {\gamma_{\text{w}}} (p_u,\mathbf{x};\xi) \geq \gamma_{\min} \}$ and the optimization problem
\begin{equation} \label{subproblem:Double_Lower_Bound}
\underline{\underline{\operatorname{EE}}}_{\text{w}} (p_u)  \coloneqq  \max_{\mathbf{x} \in \mathcal{X}} \,\! {{\operatorname{EE}_{\text{w}}}(p_u,\mathbf{x};\xi)} = \max_{\mathbf{x} \in \mathcal{X}} {F(0; \mathbf{x})}  .
\end{equation}
In step 12 Algorithm~\ref{algorithm:AO} is initialized with $p_{\text{opt}}^{(0)} = p_u$, hence  
\begin{equation} \label{equation:LB_inequalities}
\underline{\operatorname{EE}}_{\text{w}} (p_{\ell},p_u) \geq \underline{\underline{\operatorname{EE}}}_{\text{w}} (p_u)  .
\end{equation}
By combining \eqref{subproblem:Upper_Bound} and \eqref{equation:Taylor_formula}--\eqref{equation:LB_inequalities}, we can write   
\begin{equation}
\begin{split}
\overline{\operatorname{EE}}_{\text{w}} (p_{\ell},p_u) & = \max_{\mathbf{x} \in \mathcal{X}} {F(\Delta p; \mathbf{x})} \\
& \leq \max_{\mathbf{x} \in \mathcal{X}} {F(0; \mathbf{x})} + \max_{\mathbf{x} \in \mathcal{X}} {R(\Delta p; \mathbf{x})} \\
& \leq \underline{\operatorname{EE}}_{\text{w}} (p_{\ell},p_u) + M |\Delta p|   .
\end{split}
\end{equation}
Therefore, we have $0 \leq \overline{\operatorname{EE}}_{\text{w}} (p_{\ell},p_u) - \underline{\operatorname{EE}}_{\text{w}} (p_{\ell},p_u) \leq M |\Delta p|$. Finally, by taking the limit as $\Delta p \to 0$, we obtain~\eqref{equation:BnB_convergence_sufficient_condition}.

\end{document}